\documentclass[
  aps, 
  prx, 
  reprint,
  showkeys,
  nobibnotes,
]{revtex4-2}

\usepackage
  [final]{changes}
\definecolor{addedcolor}
  {rgb}{0.0, 0.7, 0.7}
\definecolor{deletedcolor}
  {rgb}{0.9, 0.4, 0.0}
\setaddedmarkup
  {\textcolor{addedcolor}{#1}}
\setdeletedmarkup
  {\textcolor{deletedcolor}{\sout{#1}}}

\usepackage{enumitem}
\setlist[enumerate,1]{label=\textbf{(\roman*)}}
\setlist{
  leftmargin=.65cm,
}

\usepackage{lmodern}
\usepackage{anyfontsize}
    
\usepackage{multirow}
\usepackage{hhline}

\usepackage{tabularray}
\UseTblrLibrary{booktabs}

\usepackage{adjustbox}
\usepackage{tcolorbox}

\newtcolorbox{standout}{
  colback=gray!15,
  boxrule=0pt,
  left=.3cm,
  right=.3cm,
  top=.18cm,
  bottom=.18cm,
  boxsep=0pt
}

\usepackage{xcolor}

\usepackage{mathtools}
\usepackage{amssymb}
\usepackage{amsthm}
\usepackage{xfrac}
\usepackage{stmaryrd} 
\usepackage{scalerel} 
\usepackage{stackengine} 

 \newcommand{\bracket}[3]{%
  \stretchleftright
    {#1}
    {%
      \ensurestackMath{\addstackgap[1pt]{#2}}%
      \vrule width 0pt depth 2pt height 0pt
    }
    {#3}%
} 
\newcommand{\scaledbracket}[3]{%
  \ThisStyle{%
    \stretchleftright
      {#1}
      {
        \ensurestackMath{\addstackgap[1pt]{\SavedStyle #2}}%
        \vrule width 0pt depth 1.5pt height 0pt
      }
      {#3}%
  }%
}
\newcommand{\bracketmid}[4]{%
  \stretchleftright{#1}{%
    \ensurestackMath{%
      \addstackgap[2pt]{#2}%
      \,\stretchrel*{|}{\addstackgap[2pt]{#2#3}}\,%
      \addstackgap[2pt]{#3}%
    }%
  }{#4}%
}

\theoremstyle{plain}

\theoremstyle{definition}

\theoremstyle{remark}

\usepackage[
  colorlinks=true,
  linkcolor=darkgreen,
  citecolor=darkgreen,
  urlcolor=darkgreen
]{hyperref}
\usepackage{xurl} 

\usepackage{cleveref}
\crefname{equation}{}{}
\crefname{section}{\S}{\S\S}
\crefname{subsection}{\S}{\S\S}
\crefname{subsubsection}{\S}{\S\S}
\crefname{definition}{Def.}{Defs.}
\crefname{theorem}{Thm.}{Thms.}
\crefname{corollary}{Cor.}{Cors.}
\crefname{lemma}{Lem.}{Lems.}
\crefname{proposition}{Prop.}{Props.}
\crefname{remark}{Rem.}{Rems.}
\crefname{notation}{Ntn.}{Ntns.}
\crefname{fact}{Fact}{Fact}
\crefname{example}{Ex.}{Exs.}
\crefname{figure}{Fig.}{Figs.}
\crefname{table}{Tab.}{Tabs.}
\crefname{footnote}{ftn.}{ftns.}
\Crefname{footnote}{Ftn.}{Ftns.}

\usepackage{tikz}
\usetikzlibrary{
  cd,
  calc,
  arrows.meta,
  backgrounds,
  decorations,
  decorations.pathmorphing, 
}

\tikzcdset{
  arrow style=tikz, 
  diagrams={
    trim left=
    {([xshift=5pt]current bounding box.west)},
    trim right=
    {([xshift=-5pt]current bounding box.east)},
    baseline=
    {([yshift=-2pt]current bounding box.center)},
    >={
      Computer Modern Rightarrow[
        length=4pt, width=4pt
      ]
    },
    hook/.style={
      {Hooks[right, length=2pt, width=8pt]}->
    },
    hook'/.style={
      {Hooks[left, length=2pt, width=8pt]}->
    },
    shorten=0pt,
  }
}

\definecolor{darkblue}{rgb}{0.05,0.25,0.65}
\definecolor{darkgreen}{RGB}{20,140,10}
\definecolor{lightgray}{rgb}{0.9,0.9,0.9}
\definecolor{darkorange}{RGB}{200,100,5}
\definecolor{darkyellow}{rgb}{.91,.91,0}
\definecolor{lightolive}{RGB}{225, 220, 185}

\let\originalsslash\sslash
\renewcommand{\sslash}{\mathord{\originalsslash}}

\makeatletter
\newcommand{\cpt}{\mathpalette\cpt@inner\relax}
\newcommand{\cpt@inner}[2]{%
  \scalebox{0.5}[0.9]{$#1\cup$}
  #1\{\infty\}
}
\makeatother

\newcommand{\grayunderbrace}[2]{\mathcolor{gray}{\underbrace{\mathcolor{black}{#1}}}_{\mathcolor{gray}{#2}}}

\tikzset{
  snake left/.style={
    rounded corners,
    to path={
      let \p1 = (\tikztostart.east),
          \p2 = (\tikztotarget.west),
          \p3 = ($(\p1)!0.5!(\p2)$),
          \n1 = {8pt} 
      in
      (\p1)
      -- (\x1 + \n1, \y1)
      -- (\x1 + \n1, \y3)
      -- (\x2 - \n1, \y3) \tikztonodes
      -- (\x2 - \n1, \y2)
      -- (\p2)
    }
  }
}

\tikzset{
  uphordown/.style={
    rounded corners,
    to path={
      let \p1 = (\tikztostart.north),
          \p2 = (\tikztotarget.north),
          \n1 = {max(\y1,\y2) + 8pt}
      in
      (\p1)
      -- (\x1, \n1)
      -- (\x2, \n1) \tikztonodes 
      -- (\p2)
    }
  }
}

\tikzset{
  downhorup/.style={
    rounded corners,
    to path={
      let \p1 = (\tikztostart.south),
          \p2 = (\tikztotarget.south),
          \n1 = {min(\y1,\y2) - 8pt}
      in
      (\p1)
      -- (\x1, \n1)
      -- (\x2, \n1) \tikztonodes 
      -- (\p2)
    }
  }
}

\tikzset{
  rightvertleft/.style={
    rounded corners,
    to path={
      let \p1 = (\tikztostart.east),
          \p2 = (\tikztotarget.east),
          \n1 = {max(\x1,\x2) + 8pt}
      in
      (\p1)
      -- (\n1, \y1)
      -- (\n1, \y2) \tikztonodes 
      -- (\p2)
    }
  }
}

\tikzset{
  leftvertright/.style={
    rounded corners,
    to path={
      let \p1 = (\tikztostart.west),
          \p2 = (\tikztotarget.west),
          \n1 = {min(\x1,\x2) - 8pt}
      in
      (\p1)
      -- (\n1, \y1)
      -- (\n1, \y2) \tikztonodes 
      -- (\p2)
    }
  }
}

\newcommand{\inlinetikzcd}[1]{\begin{tikzcd}[sep=small, ampersand replacement=\&]#1\end{tikzcd}}

\newcommand{\defneq}{\equiv}

\newcommand{\HilbertSpace}{%
  \mathcal{H}%
}

\newcommand{\Aeff}{A}

\begin{document}

\title
{
  Non-Perturbative SDiff Covariance of
  Fractional Quantum Hall Excitations
}


\author{Hisham Sati}
\email{hsati@nyu.edu}

\author{Urs Schreiber}
\email{us13@nyu.edu}

\affiliation{Mathematics Program and Center for Quantum and Topological Systems, New York University Abu Dhabi, UAE}



\date{\today}


\begin{abstract}
Collective excitations of Fractional Quantum Hall (FQH) liquids at long wavelengths are thought to be of a generally covariant geometric nature, governed by area-preserving diffeomorphisms ($\mathrm{SDiff}$). But current analyses rely solely on the corresponding perturbative $w_\infty$ Lie algebra. We argue this is insufficient: We identify a non-perturbative construction of the effective Maxwell-Chern-Simons quantum field theory which carries unitary $\mathrm{SDiff}$ equivariance. But this turns out to be non-differentiable, suggesting 
\replaced
{
underappreciated subtleties when the usual Hilbert space truncation is removed.
}
{
FQH excitation phenomenology beyond the $w_\infty$ algebra.
}
\end{abstract}

\maketitle

\deleted{[Table of Contents]}

\section{Introduction}

  Fractional quantum (anomalous) Hall systems (FQH) are noteworthy for the experimentally verified topological order they exhibit in their gapped ground states, making them candidates for future platforms of much-anticipated intrinsically stabilized quantum computing hardware. Yet more remarkably, above these topological ground states, FQH liquids are now expected to exhibit collective excitations which, at long wavelengths, are effectively described by a kind of supergravity. In this setting, chiral graviton states are expected to exhibit general covariance under the group $\mathrm{SDiff}$ of area-preserving diffeomorphisms of the 2-dimensional FQH liquid, accompanied by gravitino-like superpartner excitations promoting this to a supergroup. First signatures of both excitations have been experimentally observed. 

  These fascinating recent developments (references below) call for a more thorough and rigorous theoretical understanding of the nature of FQH excitations and their $\mathrm{SDiff}$ covariance.

\subsection{FQH Systems}

Fractional Quantum Hall Systems (FQH, cf. \cite{Stormer99,PapicBalram2024}) are ultracold, effectively 2-dimensional electron gases penetrated by a transverse magnetic field which is so strong and yet so fine-tuned that there is some integer (or, generally, rational) number $K$ of magnetic flux quanta per electron. Relatedly, fractional quantum \emph{anomalous} Hall Systems 
(cf. \cite{ju2024fractional,moralesduran2024fractionalized,zhao2025exploring}) are crystalline topological insulators with ``flat'' topological electron bands (\emph{fractional Chern insulators}), where the role of the external magnetic flux is instead played by an intrinsic crystalline property: the \emph{Berry curvature} of the valence band over the Brillouin torus of crystal momenta. 

Both kinds of systems are thought to have isomorphic effective quantum properties under a \emph{duality} (cf. \cite[Fig. 3]{SS25-FQAH}) which exchanges electron positions and magnetic flux density with electron crystal momenta and \emph{Berry curvature}, respectively; and we will use ``FQH'' to refer to either case. The non-anomalous real-space version is experimentally well-established, but the  ``anomalous'' crystalline version (experimentally realized just recently) has the tantalizing advantage for engineering applications that the required low temperatures are much less extreme and the extreme external magnetic fields are unnecessary.

While the eponymous signature of FQH systems is their quantized \emph{Hall resistivity}, the interest in these systems as quantum materials is that they are \emph{topological phases of matter} whose degenerate gapped ground states are \emph{topologically ordered} (cf. \cite{Stanescu2020}), meaning essentially that they follow laws of \emph{topological quantum field theory} (TQFT, cf. \cite{Simon2023}). It is this property and the fact that its experimental signatures are consistently being seen (starting with \cite{Nakamura2020}, recent pointers in \cite{Veillon2024}) which makes FQH systems a rare instance of candidate platforms for future topological quantum computing technology (cf. \cite{DasSarmaFreedmanNayak2005}), especially in their more amenable ``anomalous'' version, cf. \cite{Urton2023}.

But even if it is the ground states of FQH systems that carry this TQFT intrigue, one needs to understand the gapped excitations above these ground states, if only to control the topological ground-state phase under realistic non-ideal conditions.

\subsection{FQH Excitations}
\label{OnFQHExcitations}

The lowest collective excitation modes of an FQH liquid are understood to be bosonic density waves known as the \emph{magneto-roton} or \emph{GMP mode} due to \cite{GMP1986}, and a fermionic analog present at certain ``non-abelian'' filling fractions, known simply as the \emph{neutral fermion mode} \cite{YHPH2012}. It has been argued \cite{YHPH2012,Haldane2013} that at long wavelengths (hence in the \emph{infrared} where the microscopic details of the system are not resolved and an \emph{effective field theory} description may apply), these modes look like a massive spin-2 \emph{chiral graviton} (cf. \cite{Wang2023}, akin to the expected but elusive quantum of gravity in high energy physics) and a \emph{superpartner} spin-$\sfrac{3}{2}$ \emph{gravitino} (akin to the yet more elusive fermionic quantum of gravity present in hypothetical supersymmetric versions of gravity, SuGra). 
The actual supersymmetry between these two kinds of modes can be made precise \cite{GromovMartinecRyu2020} and has been experimentally observed \cite{Pu2023Signatures}, prompting attempts to develop an effective supergravity description of FQH systems \cite{NPBG2023}.

In fact, the symmetry algebra generated by the GMP operators \cite[(4.13)]{GMP1986} expected to create the magneto-roton mode from the ground state (an \emph{FFZ} or  \emph{$W_\infty$-algebra} \cite{FFZ1989,CTZ1994}) transitions at long wavelength into the Lie algebra of \emph{area-preserving diffeomorphisms} (APDs), hence into a form of the \emph{general covariance} which is the hallmark of gravitational theories, but reflecting the characteristic incompressibility of the FQH liquid. This suggests that effective FQH excitation theory should be an APD gauge theory \cite{IKS1992,CTZ1993,KarabaliNair2004,DMNXS2022,WangYang2023,Du2025}!

We highlight that, remarkably, APDs are in fact the symmetries seen on the \emph{membrane} (M2-brane) probe of 11D supergravity \cite{Floratos1988,dWHH1988,dWMH1990}, and that it is generally on $p$-dimensional such brane probes of SuGra that volume-preserving diffeomorphism symmetry appears \cite{BSTT1990,BandosTownsend2008,HIMS2008}. This will be relevant for the \emph{geometric engineering} (cf. \cite{nLab:GeometricEngineering}) of FQH systems on branes, as in \cite{SS25-Seifert,SS25-Srni}.

\subsection{Open Problems}

A comprehensive understanding of FQH excitations, therefore, appears to hinge on a good understanding of the action of APD symmetry on quantum states of FQH liquids. This, however, seems to have been lacking:

For a surface $\Sigma^2$, the APD symmetry group --  denoted 
\begin{equation}
  \label{TheSDiffGroup}
  \substack{
    \text{\color{gray}APD}
    \\
    \text{\color{gray}symmetry}
  }
  \;\;
  \mathrm{SDiff}\bracket({\Sigma^2})
  \subset
  \mathrm{Diff}\bracket({\Sigma^2})
  \;\;
  \substack{
    \text{\color{gray}general}
    \\
    \text{\color{gray}covariance}
  }
\end{equation}
in the mathematical literature (the \emph{special} subgroup of the full diffeomorphism group) --  is an infinite-dimensional Fr{\'e}chet-Lie group, and as such its representation theory is subtle (cf. \cite{Milnor1984,Ismagilov1996}). In particular, representations of such groups on Hilbert spaces $\HilbertSpace$ 
\begin{equation}
  \label{TheRepresentation}
  \begin{tikzcd}
    U_{(-)}
    :
    \mathrm{SDiff}\bracket({\Sigma^2})
    \ar[r]
    &
    \mathrm{U}\bracket({\HilbertSpace})
  \end{tikzcd}
\end{equation}
are generically \emph{not differentiable} (and we will see in \cref{OnGlobalCovariance} that this is the case here!).
This means that the study of its ``perturbation theory'' in the guise of the corresponding Lie algebra 
\begin{equation}
  \label{TheLieAlgebra}
  w_\infty\bracket({\Sigma^2})
  := \mathrm{Lie}\bracket({\mathrm{SDiff}\bracket({\Sigma^2})})
\end{equation}
may fail to correctly represent the infrared FQH excitations.

Indeed, this has recently been noticed in special cases: The single GMP mode already at long wavelengths, understood as the action of such Lie algebra elements on a vacuum state (cf. \cite[(4)]{GromovMartinecRyu2020} going back to \cite{GMP1986}),
\begin{equation}
  \label{TheGMPStates}
  \overline{\rho}
  \,
  \vert 0\rangle
  \,,
  \;\;\;
  \overline{\rho} \in w_\infty
  \mathrlap{\,,}
\end{equation}
\emph{fails} even to be an accurate approximation of the actual excited FQH states, at least
\begin{enumerate}
\item for filling fractions near $\nu = \sfrac{1}{4}$ \cite{NHRSTY2022},

\item for  $\nu = \sfrac{n}{(2pn \pm 1)}$ with $\vert n \vert, p > 1$ \cite{BLGP2022,BalramSreejithJain2024},

\item for spinful (non-polarized) electron liquids \cite{DoraBalram2025}. 
\end{enumerate}
In fact, according to \cite{BLGP2022,BalramSreejithJain2024} these failures are intrinsically due to an internal \emph{parton}/\emph{exciton} structure of the GMP mode, so that even where the GMP mode is an accurate approximation (such as at $\nu = \sfrac{n}{(2n\pm 1)}$ \cite{BalramSreejithJain2024}) it is not an exact quantum state of the FQH system.  

But this raises the conceptual question of what the APD symmetry action \cref{TheRepresentation} on FQH systems actually is, if not induced by a $w_\infty$ Lie action \cref{TheGMPStates} taking quantum states to quantum states, as commonly advertised!

Here, we aim to clarify at least a fundamental aspect of this problem. Doing so might be particularly interesting in that it introduces constructive QFT methods to the topic that may not have been applied to FQH systems before, and whose conclusions may lie outside of what can be achieved with more traditional methods.

\section{Global APD Covariance}
\label{OnGlobalCovariance}

Namely, we turn to methods of \emph{constructive quantum field theory} (cf. \cite{GlimmJaffe1987}) for the rigorous construction of non-perturbative quantum field theories, here applied to the effective field theory of FQH excitations.

\subsection{Effective MCS Theory}

First we recall (cf. \cite[(11.41)]{Fradkin2013}\cite[\S 10]{Fradkin2024}\cite[\S 2.3]{Willsher2020}) that at long wavelengths FQH liquids are effectively described, one way or another, by Maxwell-Chern-Simons theory (MCS, general review in \cite[\S 2.2]{Dunne1998}\cite[\S 2.2.3]{Moore2019}) of an effective gauge potential 1-form $\Aeff$ 
with Lagrangian density locally of the form (cf. \cite[(1)]{KarabaliKimNair2000}):
\begin{equation}
  \label{TheMCSLagrangianDensity}
  L(\Aeff)
  :=
  \grayunderbrace{
    \frac{\tilde m}{T}
    \mathrm{d} \Aeff
    \wedge 
    \star_3 
    \mathrm{d} \Aeff
  }{
    \mathclap{\text{Maxwell}}
  }
  \,+\,
  \grayunderbrace{
    \frac{k}{4\pi}
    \Aeff \wedge \mathrm{d} \Aeff
  }{
    \mathclap{\text{Chern-Simons}}
  }
  \mathrlap{\,,}
\end{equation}
where we use the standard physics convention for the flux quantization (cf. \cite[\S 2.1]{SS25-Flux}) of the flux density $F$ (locally given by $\mathrm{d}A$), normalizing it such that $\tfrac{1}{2\pi}F$ has integral periods.

The constants in \cref{TheMCSLagrangianDensity} are:
\begin{itemize}

\item
$\sqrt{T/\tilde m}$, the \emph{coupling constant} expressed relative to the mass gap $\tilde m$ of the theory (for this notation cf. \cref{MatchingConcepts}),
\item
$k \in \mathbb{Z}$, the \emph{Chern-Simons level} (in its divided integer form appropriate for spin-TQFT, cf. \cite[\S 2]{SS25-WilsonLoops}),  being the inverse  of the \emph{filling fraction} $\nu = \sfrac{1}{k}$ of the FQH system (cf. \cite[(2) \& (8)]{SS25-FQH}). 
\end{itemize}

In the \emph{strong coupling limit} --- $T \to \infty$ at fixed $k, \tilde{m}$ --- the kinetic Maxwell term in \cref{TheMCSLagrangianDensity} becomes negligible and nominally only the topological Chern-Simons (CS) interaction remains (cf. \cite[p. 20]{Dunne1998}). This is traditionally used to describe the topological ground states of FQH systems (going back to \cite{Zhang1992,Wen1995}, review in \cite{Willsher2020}).

But here we are concerned with the case of \emph{finite} coupling, $T < \infty$, where the kinetic energy plays a role, and FQH excitations above the topological ground state are supposed to appear. Incidentally, it has been argued (\cite[\S IX]{Haller1996}) that some excitations do survive even in the limit, so that MCS for $T \to \infty$ is still richer than pure $\mathrm{CS}$, and we will find this borne out in rigorous detail (cf. \cref{OnCSVacuaInMCSTheory}).

\subsection{Its Quantum States}
\label{OnTheQuantumStates}

For the canonical quantization (cf. \cite{BlaschkeGieres2021}\cite[\S 19]{HenneauxTeitelboim1992}) of the Maxwell-Chern-Simons Lagrangian \cref{TheMCSLagrangianDensity}, 
we consider a spacetime of the globally hyperbolic form
\begin{equation}
  \label{TheSpacetimeDecomposition}
  X^{1,2}
  \simeq
  \mathbb{R}^{1,0}
  \times
  \Sigma^2
  \mathrlap{,}
\end{equation}
where $\Sigma^2$ denotes a closed orientable surface to be understood as the region inhabited by the effectively 2-dimensional FQH electron liquid.

Focusing on the topologically trivial gauge field sector for the time being (which is the only case that occurs for $\Sigma^2 = S^2$, but not for higher genus surfaces), the configuration space of the theory is that of \emph{gauge-potential} 1-forms 
\begin{equation}
  \label{TheGaugePotentials}
  A \in \Omega^1_{\mathrm{dR}}\bracket({
    \Sigma^2 ; \mathrm{i}\mathbb{R}
  })
  \mathrlap{\,.}
\end{equation}
Classically, these are \emph{smooth} differential forms, but quantum mechanically (``in the path integral''), a key subtlety is identifying the appropriate stochastic generalization class, which we turn to in a moment.

However, first note that the CS-term in \cref{TheMCSLagrangianDensity} ``twists'' the \emph{Gauss law} constraint (cf. \cite[(5.8)]{BlaschkeGieres2021}\cite[(19.8)]{HenneauxTeitelboim1992}). This results in the quantum 
states of the theory being (``wave'')functions $\Psi$ of the gauge potentials \cref{TheGaugePotentials} which are not quite invariant under gauge transformations $\Aeff \mapsto \Aeff + \mathrm{d}\xi$ but instead pick up a complex phase, as follows (cf. \cite[(94)]{Dunne1998}\cite[(7)]{KarabaliKimNair2000}):
\begin{equation}
  \label{GaussLawOnWavefunctions}
  \Psi\bracket({\Aeff + \mathrm{d}\xi})
  =
  e^{
    \tfrac
      {k \mathrm{i}}
      {4\pi}
    \int_{\Sigma^2}
    \Aeff \wedge \mathrm{d}\xi
  }
  \,
  \Psi\bracket({\Aeff})
  \mathrlap{\,.}
\end{equation}
Notice that this transformation property is independent of the coupling constant $\sqrt{T}$ and, in particular, is the same for plain Chern-Simons theory (for which cf. \cite[(2.191)]{Moore2019}). Hence it applies to the effective description of FQH systems both generally and in the topological limit.

Therefore, a rigorous construction of the Hilbert space $\HilbertSpace$ of quantum states of MCS theory \cref{TheMCSLagrangianDensity}, schematically
\begin{equation}
  \label{TheHilbertSpaceSchematically}
  \HilbertSpace
  =
  \bracket\{
    {\text{wavefunctions }\Psi\text{ satisfying \cref{GaussLawOnWavefunctions}}}
    \}
  \mathrlap{\,,}
\end{equation}
requires, for its inner product $\langle - \vert - \rangle$, the construction of a suitable Gaussian (``path integral'') measure on a space of Hermitian squares of wavefunctions $\Psi$ as in \cref{GaussLawOnWavefunctions} by integration over a suitable space of gauge potentials \cref{TheGaugePotentials}, such that the result tends to the Hilbert space of plain Chern-Simons theory in the limit $T \to \infty$.   

This may seem demanding, but our main observation now is (detailed in \cref{OnCanonicalMCSTheory}) that a rigorous such construction may be identified from a careful utilization of \cite{Pickrell2000} (based on \cite{Pickrell1996}, both done with a formal purpose, not motivated by our physical context), which proceeds by functional integration methods of non-perturbative constructive quantum field theory (cf. \cite{GlimmJaffe1987}) as follows.

Subject to (heavy) analytic details, the Gaussian measure is (cf. \cite[p. 64]{Pickrell2022}) a renormalized form of the expected exponential kinetic Maxwell Lagrangian in \cref{TheMCSLagrangianDensity} restricted to the spatial leaf $\Sigma^2$; making precise this informal spatial ``path integral'' expression:
\begin{equation}
  \label{TheInnerProduct}
  \bracketmid\langle{\Psi_1}{\Psi_2}\rangle
  \!=\!\!\!
  \int_{\mathrlap{[\Aeff]}}
  \,
  \overline{\Psi}_{\!1}(\Aeff)
  \Psi_{\!2}(\Aeff)
  \,
  e^{\bracket({
    \tfrac{-1}{2T}
    \!\!
    \int_{\Sigma^2}
    \!
    \mathrm{d}\Aeff
    \wedge \star_2
    \mathrm{d}\Aeff
  })}
  \mathcal{D}[A]
\end{equation}
over gauge equivalence classes $[A]$ of field configurations. 

This follows from equations (2.3)-(2.33) of \cite{Pickrell2000}, noting that the starting-point equations (2.1) \& (4.7)
in op. cit. (where the latter has the proper prefactor) are just our Gauss law \cref{GaussLawOnWavefunctions}; cf. also \cite[(4)]{KarabaliKimNair2000}.
\added{In \S\ref{OnCanonicalMCSTheory} we relate \eqref{TheInnerProduct} to the expected expression in the physics literature.}

Little further is currently known about this exact non-perturbative Hilbert space \cref{TheInnerProduct} (for lack of investigation) --- except for the following remarkable result, which appears to rigorously bring out the expectations about FQH excitations from \cref{OnFQHExcitations}.

\subsection{The APD Covariance}
\label{OnTheAPDCovariance}

The main result now is that\added{, over a closed surface $\Sigma^2$}:
\begin{enumerate}
\item
the canonical action of $\mathrm{SDiff}\bracket({\Sigma^2})$ \cref{TheSDiffGroup} (via pullback of gauge potentials $\Aeff$) does induce a continuous unitary representation $U_{(-)}$ \cref{TheRepresentation} on the MCS quantum states \cref{TheHilbertSpaceSchematically}, by
\cite[\S 2]{Pickrell2000};
\item this action is \emph{not differentiable}, by \cite[Rem. 2.36]{Pickrell2000};
\item these statements continue to hold in the \emph{strong coupling limit} $T \to \infty$ where the Chern-Simons space of states is recovered, by \cite[p. 179 \& \S 3]{Pickrell2000}.
\end{enumerate}

\added{
To put this in perspective, notice that all previous discussions of the algebra of FQH excitations on the sphere (recently in \cite{DoraBalram2025,He2025,EckWang2026}) consider truncation to bounded angular momentum, where the algebra becomes finite-dimensional. While (as observed for membrane dynamics long ago, \cite[p. 563]{dWHH1988}) the Lie algebra of area-preserving diffeomorphisms is recovered as the truncation is lifted (cf. \cite{EckWang2026}), the analogous limit \emph{fails dramatically} at the level of groups (as has been pointed out long ago in \cite{swain2004limiting,*swain2004topology}), rendering the truncation non-perturbatively inconsistent.
}

\added{
Concretely, \cite[Rem. 2.36]{Pickrell2000} shows that the failure of the would-be Lie derivative to exist is a superficial divergence of the norm of the would-be quantum state \eqref{TheGMPStates}. That norm is proportional to the \emph{static structure factor} of the FQH liquid (by \cite[(2.7)]{GMP1986}), whose definition may hence need more attention in the untruncated effective theory.
}

\deleted{
Recall here that the strong coupling limit $T \to \infty$ is the passage from the dynamical Maxwell-Chern-Simons theory to its topological Chern-Simons sector (off-shell).
}
\deleted{
For discussion of FQH excitations, we are interested in finite values of $T$; while in the limit, the excited states should disappear and only the topological ground states remain, on which the $\mathrm{SDiff}$-action ought to factor through the modular group of connected components.}

\section{Conclusion}
\label{Conclusion}

Remarkably, the result of \cref{OnTheAPDCovariance} suggests that:
\begin{standout}
  The APD symmetry/excitation spectrum of FQH liquids is realized on the genuine Hilbert space of FQH quantum states at the level of the full topological symmetry group $\mathrm{SDiff}$,
  but \emph{not} at the $w_\infty$ Lie algebraic level of projected density operators traditionally assumed. 
\end{standout}

This would mean that the traditional schematic expression for excited FQH states (cf. \cite[(4)]{GromovMartinecRyu2020} going back to \cite{GMP1986}),
\begin{equation}
  \label{TraditionalGMPModeExpression}
  \vert \phi_{\mathbf{k}} \rangle
  :=
  \overline{\rho}_{\mathbf{k}}
  \vert \Psi_0 \rangle
  \mathrlap{\,,}
\end{equation}
is --- at least in the long-wavelength limit where the \emph{projected density operator} $\overline{\rho}_{\mathbf{k}}$ may be regarded as a would-be representation of an element of the $w_\infty$ Lie algebra \cref{TheLieAlgebra} --- not actually a quantum state in the Hilbert space $\HilbertSpace$ (\cref{OnTheQuantumStates}) of the FQH liquid, in general, even where it is an excellent approximation to one 
\added{in a truncated Hilbert space}. 

Instead, the facts in \cref{OnTheAPDCovariance} say that what would exist as actual FQH quantum states are expressions of the form
\begin{equation}
  \label{FiniteGMPModeProposal}
  \vert \phi_{\kappa} \rangle
  :=
  \tfrac{1}{\epsilon}
  \bracket({U_\kappa - \mathrm{id}})
  \vert \Psi_0 \rangle
  \mathrlap{\,,}
\end{equation}
for $\epsilon$ a normalization factor and $\kappa \in \mathrm{SDiff}\bracket({\Sigma^2})$ a \emph{finite} area-preserving diffeomorphism \cref{TheSDiffGroup}, with $U_{\kappa}$ its unitary operator representation \cref{TheRepresentation} according to \cref{OnTheAPDCovariance}.

The traditional expression \cref{TraditionalGMPModeExpression} implicitly requires that suitable limits of \cref{FiniteGMPModeProposal} of the form $\inlinetikzcd{\kappa \ar[r] \& \mathrm{e}}$, $\inlinetikzcd{\epsilon \ar[r] \& 0}$ exist, but this is generally not the case non-perturbatively, according to \cref{OnTheAPDCovariance}. 

It would remain to be clarified how exactly this discrepancy arises. Currently, we have two opposite approaches to determining the quantum states of excited FQH systems:
\begin{enumerate}
\item
Traditionally: 
explicit Laughlin-type \emph{Ans{\"a}tze} (educated guesses, going back to \cite{Laughlin1983}) for FQH ground states $\vert \Psi_0\rangle$, argued to be accurate (as opposed to exact) in the purely topological sector (cf. \cite[\S 3.1]{Simon2020}), and then acted on by projected density operators \cref{TraditionalGMPModeExpression} argued to, in turn, produce accurate (as opposed to exact) excited quantum states from the approximate topological ground states.

\item
Here: abstract characterization of the exact non-topological FQH states \added{in the long-wavelength limit}, according to \cref{OnTheQuantumStates}.
\end{enumerate}

Apparently, there is a gap between these two approaches that remains to be filled. Meanwhile, it appears that the effective description of FQH excitations via (super-)APD covariant brane/gravity field theories (as envisioned in \cite{IKS1992,CTZ1993,DMNXS2022,WangYang2023,Du2025}) may need to shift attention beyond the perturbative $w_\infty$ symmetry (which does not actually exist exactly) to the full non-perturbative $\mathrm{SDiff}$ symmetry established in \cref{OnGlobalCovariance}.

Here it appears suggestive that such $\mathrm{SDiff}$ symmetry is naturally expected not in effective gravity theories (which of course should be $\mathrm{Diff}$-covariant, instead) but on \emph{brane probes} of gravitational targets (by \cite{BSTT1990}), specifically (by \cite{BandosTownsend2008}) on \emph{M-branes} (cf. \cite{GSS25-M5}\added{, for which analytic subtleties in the removal of the regulator have been pointed out long ago \cite{swain2004limiting,*swain2004topology}}), and 
for which we have recently shown (\cite{SS25-Seifert,SS25-Srni}) how their topological sector coincides in fine detail with the topological order of FQH liquids (based on \cite{SS25-AbelianAnyons,SS25-FQH}).

\appendix

\section{Supplementary Material}

\subsection{Canonical MCS Theory}
\label{OnCanonicalMCSTheory}

In \cref{OnTheQuantumStates} we claim that the Hilbert space constructed by Pickrell (2000) in \cite[\S 2,3]{Pickrell2000} is that of quantum states of Maxwell-Chern-Simons theory at any finite coupling. Here, we indicate how to see that this is indeed the case.

To that end, we invoke the analysis of canonical quantization of 3D Yang-Mills-Chern-Simons theory due 
to Karabali, Kim \& Nair (KKN) \cite{KarabaliKimNair2000} (following \cite{KarabaliNair1996}) from the same year --- which was given independently and has not before been related to Pickrell's result. 
It is then a matter of carefully matching the notation and conventions between these two (groups of) authors. The result is shown in \cref{MatchingConcepts} (where we straightforwardly specialize the KKN expressions to the abelian case of interest here).

\begin{table}[htb]
\caption{
  \label{MatchingConcepts}
  Notations and conventions
  in canonical MCS theory, for comparison in \cref{OnCanonicalMCSTheory}. Here, in specializing the discussion of YMCS theory of \cite{KarabaliKimNair2000} to the abelian case of MCS theory, we use $\mathrm{Tr}(B^2) = \tfrac{1}{2} B^2$ (cf. \cite[p. 2]{KarabaliKimNair2000}) and $c_A = 0$ (the quadratic Casimir of the Lie algebra, cf. \cite[p. 4]{KarabaliNair1996}). All integrals are over the surface $\Sigma^2$
  \cref{TheSpacetimeDecomposition}
  against the volume form.
}
\begin{tblr}{
  colspec = {c|c|c},
  colsep = 2pt,
  row{even} = {bg=gray!15}
}
  \toprule
  \textbf{Concept}
  &
  KKN 2000
  \cite{KarabaliKimNair2000}
  &
  Pickrell 2000
  \cite{Pickrell2000}
  \\
  \midrule
  CS Level
  & 
  \begin{tabular}{c}
  $k$
  \\
  \footnotesize
  cf. (1) there
  \end{tabular}
  &
  \begin{tabular}{c}
    $1,2$
    \\
    \footnotesize
    impl. in (4.7), (4.2)
  \end{tabular}
  \\
  3D Coupling
  &
  \begin{tabular}{c}
  $e$
  \\
  \footnotesize
  cf. (1)
  \end{tabular}
  &
  ---
  \\
  mass gap
  &
  \begin{tabular}{c}
  $\tilde m = k e^2 / 4\pi$
  \\
  \footnotesize
  below (38)
  \end{tabular}
  &
  ---
  \\
  2D Coupling 
  &
  \begin{tabular}{c}
  $g = \sqrt{\tilde m} \, e$
  \\
  \footnotesize
  below (39)
  \end{tabular}
  &
  \begin{tabular}{c}
  $\sqrt{T}$
  \\
  \footnotesize
  (2.3)
  \end{tabular}
  \\
  \begin{tabular}{c}
  YM measure
  \\
  {\footnotesize(unrenormalized)}
  \end{tabular}
  &
  \begin{tabular}{c}
  $
  \underbrace{
  e^{
    - \tfrac{1}{2 \tilde m e^2} 
    \raisebox{1pt}{$\scriptstyle\int$}\! 
    B^2
  }
  }_{ \vert \Phi_0 \vert^2 }
  d\mu
  $
  \\
  \footnotesize
  (37) \& below (39)
  \end{tabular}
  &
  \begin{tabular}{c}
  $
    \tfrac{1}{\mathcal{Z}}
    e^{
     - \tfrac{1}{2T}
     \raisebox{1pt}{$\scriptstyle\int$}\! 
     F^2
    }
    \mathcal{D}A
  $
  \\
  \footnotesize
  (2.3)
  \end{tabular}
  \\
  Topol. weight
  &
  \begin{tabular}{c}
  $
  e^{\mathcal{S}(H)}
  =
  \mathrm{det}(D\overline{D})
  $
  \\
  \footnotesize
  (40) \& (41)
  \end{tabular}
  &
  \begin{tabular}{c}
  $\vert \mathrm{det}\vert^2$
  \\
  \footnotesize
  (2.6) \& (2.23)
  \end{tabular}
  \\
  \midrule
  \begin{tabular}{c}
    Full measure
    \\
    {\footnotesize(unrenormalized)}
  \end{tabular}
  &
  \begin{tabular}{c}
  $
  e^{k\mathcal{S}(H)}
  \,
  e^{
    -\tfrac{1}{2 g^2} 
    \raisebox{1pt}{$\scriptstyle\int$}\! 
    B^2
  }
  d\mu
  $
  \\
  \footnotesize
  (40)
  \end{tabular}
  &
  \begin{tabular}{c}
  $
    \vert \mathrm{det} \vert^2
    \,
    e^{
      -\tfrac{1}{2T}
      \raisebox{1pt}{$\scriptstyle\int$}\! 
      F^2
    }
    \mathcal{D}A
  $
  \\
  \footnotesize
  Def. 2.30 \& (2.32)
  \end{tabular}
  \\
  \bottomrule
\end{tblr}
\end{table}

Now, the last line of \cref{MatchingConcepts} obtains the ``path integral'' measure used in \cref{TheInnerProduct} for the definition of the inner product of wavefunctions.
The table shows that, in the unrenormalized form in which the expressions are given, they match between these authors. But the fully rigorous renormalized construction of this measure is just what \cite[\S 2]{Pickrell2000} establishes --- which with \cite{KarabaliKimNair2000} is thereby identified as giving rise to the full non-perturbative Hilbert space of MCS theory, as claimed.

\subsection{CS States in MCS Theory}
\label{OnCSVacuaInMCSTheory}

At the level of Lagrangian densities \cref{TheMCSLagrangianDensity}, Maxwell-Chern-Simons theory (MCS) reduces to pure abelian Chern-Simons theory (CS) in the strong coupling limit $T \to \infty$. But a moment of reflection shows that at the level of full field theories there do remain MCS modes in the limit (cf. also the Hamiltonian analysis in \cite[\S IX]{Haller1996}).

This is simply because the phase spaces do not converge in the limit:
\begin{enumerate}
\item the phase space of MCS at any finite value of $T$ is  the space of all gauge potentials and their electric field strength canonical momenta, whence in standard real polarization the wavefunctions are functions of all gauge potentials;

\item
the phase space of CS is that of flat gauge potentials only, whence its wavefunctions depend only on the ''holomorphic half'' of these (in complex polarization). 
\end{enumerate}

This is generally a (maybe underappreciated) issue for the common argument (\cite[(11.41)]{Fradkin2013}\cite[\S 10]{Fradkin2024}\cite[\S 2.3]{Willsher2020}) that MCS theory serves as the effective field theory for FQH liquids above the topological Chern-Simons vacua.

However, we now explain how Chern-Simons states are recovered in MCS theory. This may be gleaned from \cite{Pickrell2000} under our identification from \cref{OnCanonicalMCSTheory}. 

Namely, choose any complex structure $J$ on the compact surface $\Sigma^2$ inhabited by the FQH liquid (just as one does for constructing Laughlin wavefunctions). Then Hodge theory implies (cf. \cite[(2.5), (4.12)]{Pickrell2000}) a canonical decomposition of the MCS gauge field 1-forms \cref{TheGaugePotentials} as:
\begin{equation}
  \begin{tikzcd}[row sep=-2pt, column sep=0pt]
    \Omega^1_{\mathrm{dR}}\bracket({\Sigma^2;\mathrm{i}\mathbb{R}})
    \ar[rr, "{ \sim }"]
    &&
    \overline{\partial}
    \Omega^0_{\mathrm{dR}}\bracket({
      \Sigma^2;\mathbb{C}
    })
      \oplus
    \mathrm{ker}\bracket({
      \partial_{\vert \Omega^1_{\mathrm{dR}}}
    })
    \\
    A 
      &\mapsto&
    \overline{\partial}\phi
      + 
    A^{0,1}_0
    \,,
  \end{tikzcd}
\end{equation}
where:
\begin{enumerate}
\item $A^{0,1}_0$ is the $(0,1)$-component of a flat gauge field:
\begin{equation}
  A_0 
  :=
  A^{0,1}_0 
    + 
  A^{1,0}_0
  \in
  \Omega^1_{\mathrm{dR}}\bracket({
    \Sigma^2; \mathrm{i}\mathbb{R}
  })
  \,,\;
  \mathrm{d} A_0 = 0
  \mathrlap{\,,}
\end{equation}
and hence is exactly what topological CS wavefunctions should be defined on;

\item $\phi$ is a further dynamical fluctuation about this topological component, as expected in MCS theory.
\end{enumerate}
This also shows that gauge transformations $A \mapsto A + \mathrm{d}\lambda$ shift $\phi \mapsto \phi + \lambda$ by any imaginary functions $\lambda$, whence 
gauge equivalence classes $[A]$ may simply be parameterized by taking $\phi$ to be real (cf. \cite[below (2.8)]{Pickrell2000}).

Finally, it turns out that the vacuum states of MCS theory on $\Sigma^2 \defneq T^2$ factor in their dependencies on this decomposition into topological and fluctuation modes, as follows (using \cite[Lem. 4.10]{Pickrell2000} with $ \mathrm{d}z \wedge \mathrm{d}\bar z = -2\mathrm{i} \, \mathrm{vol}$):
\begin{equation}
  \def\arraystretch{0}
  \Psi_0\bracket({\bracket[A]})
  =
  e^{
    \tfrac
      {- k}
      {4\pi}
    \raisebox{1pt}{$\scriptstyle\int$}
    \scaledbracket\vert{\nabla\phi}\vert^2
    \mathrm{vol}
  }
  \,
  \underbrace{
  e^{
    \tfrac
      {- k}
      {8\pi}
    \raisebox{1pt}{$\scriptstyle\int$}
    \scaledbracket\vert A\vert^2
    \mathrm{vol}
  }
  \,
  \Theta\bracket({
    A^{0,1}_0
  })
  }_{
    \Psi_{\mathrm{CS}}
    \scaledbracket({
      \scaledbracket[{
        A^{0,1}_0
      }]
    })
  }
  \mathrlap{\,,}
\end{equation}
where $\Theta(-)$ is a section of a holomorphic line bundle over the space of gauge equivalence classes of flat gauge fields. Here the factor $\Psi_{\mathrm{CS}}$ is just of the form of a CS-wavefunction in canonical formulation (cf. \cite[(3.14)]{ElitzurMooreSchwimmerSeiberg1989}). In conclusion, we find the MCS ground state wavefunctions to be topological CS ground states of the topological field component tensored with a Gaussian ground state of the fluctuation modes about this topological vacuum.


\begin{acknowledgments}
We thank 
A. Balram,
D. Karabali,
V. P. Nair,
and
D. Pickrell
for discussion.

This research was supported by \emph{Tamkeen UAE} under the 
\emph{NYU Abu Dhabi Research Institute grant} \texttt{CG008}.
\end{acknowledgments}

\bibliography{refs.bib}

\end{document}